\documentclass[12pt]{article}
\usepackage{authblk}
\usepackage{here}
\usepackage{xcolor}
\usepackage{amsfonts}
\usepackage{amsmath,amssymb}
\usepackage{mathrsfs}
\usepackage{bm}
     
\usepackage{graphicx}          
\usepackage{ascmac}
\usepackage{braket}
\usepackage{cite}
\usepackage[top=25truemm,bottom=30truemm,left=20truemm,right=20truemm]{geometry}

\usepackage{afterpage}
\usepackage{cases}
\usepackage{url}
\numberwithin{equation}{section}

\usepackage[breaklinks=true,linktocpage=true]{hyperref}

\baselineskip=\normalbaselineskip

\hypersetup{
  linktoc = all,
  colorlinks   = true,
  linkcolor = blue,% default:red
  anchorcolor = black,% default:black
  citecolor = blue,% default:green
  filecolor = cyan,% default:cyan
  menucolor = red,% default:red
  runcolor = cyan,% default:cyan
  urlcolor = magenta,% 
}

\newcommand{\cA}{{\mathcal A}}

\newcommand{\cC}{{\mathcal C}}
\newcommand{\cD}{{\mathcal D}}
\newcommand{\cE}{{\mathcal E}}

\newcommand{\cL}{{\mathcal L}}
\newcommand{\cM}{{\mathcal M}}

\newcommand{\cP}{{\mathcal P}}

\newcommand{\cS}{{\mathcal S}}
\newcommand{\cT}{{\mathcal T}}

\newcommand{\cV}{{\mathcal V}}
\newcommand{\cW}{{\mathcal W}}
\newcommand{\cX}{{\mathcal X}}

\newcommand{\cZ}{{\mathcal Z}}

\newcommand{\bR}{{\mathbb R}}
\newcommand{\bS}{{\mathbb S}}
\newcommand{\bZ}{{\mathbb Z}}

\newcommand{\sfI}{{\mathsf I}}
\newcommand{\sfW}{{\mathsf W}}
\newcommand{\sfV}{{\mathsf V}}
\newcommand{\sfT}{{\mathsf T}}

\newcommand{\nn}{\nonumber}

\newcommand{\wed}{\wedge}
\newcommand{\er}{\eqref}
\renewcommand{\emph}{\textit}
\newcommand{\bb}{\mathbb}
\newcommand{\fr}{\frac}
\title{{\LARGE Higher-group structure in $2n$-dimensional axion-electrodynamics}}

\author[1]{Tatsuki Nakajima}
\author[1,2]{Tadakatsu Sakai}
\author[3]{Ryo Yokokura}
\affil[1]{Department of Physics, Nagoya University}
\affil[2]{Kobayashi-Maskawa Institute for the Origin of Particles 
and the Universe, Nagoya University}
\affil[3]{
KEK Theory Center, Tsukuba 305-0801, Japan,
\par 
Research and Education Center for Natural Sciences,
Keio University, Hiyoshi 4-1-1, Yokohama, Kanagawa 223-8521, Japan}

\date{}
\begin{document}
\maketitle

\begin{abstract}
We investigate $2n$-dimensional axion 
electrodynamics for the purpose of exploring a higher-group
structure underlying it.
This is manifested as a Green-Schwarz transformation
of the background gauge fields that couple minimally to
the conserved currents.
The $n=3$ case is studied most intensively.
We derive the identities of correlation functions among
the global symmetry generators by using a gauge transformation
that maps two correlation functions with each other.
A key ingredient in this computation is given by
the Green-Schwarz transformation 
and the 't Hooft anomalies 
associated with the gauge transformation.
The algebraic structure of these results and
its physical interpretations are discussed in detail.
In particular, we find that the higher-group structure for
$n=3$ is endowed with a multi-ary operation among the symmetry
generators.

\end{abstract}

 \tableofcontents
%%%%%%%%%%%%%%%%%%%%%%%%%%%%%%%%%%%%%%%%%
\section{Introduction}

Recent developments in studies of global symmetries have led to
a deeper understanding of nonperturbartive aspects of quantum field 
theories(QFTs).

One of the most important results in this line is
the use of higher-form symmetries~\cite{Gaiotto:2014kfa}
(see also Refs.~\cite{Batista:2004sc,Pantev:2005zs,Pantev:2005wj,Nussinov:2006iva,Nussinov:2008aa,Nussinov:2009zz,Nussinov:2011mz,Banks:2010zn,Distler:2010zg,Kapustin:2013uxa,Kapustin:2014gua}).
$p$-form symmetry is characterized by
a symmetry generator whose support is given by a manifold of
codimension $p+1$, and which measures a charge carried by $p$-dimensional
extended objects.
Such symmetries have been applied to many contexts of quantum field theories, see {\it e.g.}~\cite{Gaiotto:2017yup,Gaiotto:2017tne,Tanizaki:2017qhf,Tanizaki:2017mtm,Komargodski:2017dmc,Hirono:2018fjr,Hirono:2019oup,Hidaka:2019jtv,Anber:2019nze,Misumi:2019dwq,Anber:2020xfk,Anber:2020gig,Hidaka:2020ucc,Yamamoto:2020vlk,Furusawa:2020kro,Yamamoto:2022vrh}.
It may occur that 
transformation of a certain rank induces that of a different rank
in a nontrivial manner.
Such a mathematical structure is referred to as higher-group structure.
{}For instance, 2-group structure~\cite{Baez:uc} 
involves 0-form and 1-form symmetries,
where the action of the 0-form symmetry generator on the 1-form
symmetry generator gives rise to another 1-form symmetry generator.
{}For a recent development of higher-group structure 
in the QFT context, see {\it e.g.} 
\cite{Kapustin:2013uxa,Kapustin:2013qsa,Sharpe:2015mja, Bhardwaj:2016clt,Kapustin:2017jrc,Tachikawa:2017gyf, deAlmeida:2017dhy, Benini:2018reh,Cordova:2018cvg, Delcamp:2018wlb, Wen:2018zux, Delcamp:2019fdp, Thorngren:2020aph,Cordova:2020tij,Hsin:2019fhf,Hsin:2020nts,Gukov:2020btk,Iqbal:2020lrt,Brauner:2020rtz,DeWolfe:2020uzb,Heidenreich:2020pkc,Apruzzi:2021vcu,Bhardwaj:2021wif}.
Mathematical formulation of higher-group structure is made  
by means of an extension of group to higher-category, see
\cite{Baez:2010ya} for a review.

It has become clear that axion systems in four dimensions 
serve as a QFT model 
that encodes the physical and mathematical structures of higher-group
in a simple but nontrivial manner.
In particular,
the papers \cite{Hidaka:2020iaz,Hidaka:2020izy} show that 
the massless axion and Maxwell system exhibits a 3-group
structure. 
When the axion and photon is massive, it enhances to a 4-group
structure \cite{Hidaka:2021kkf}.
The higher-group structure in axion-Yang-Mills theories is
discussed in \cite{Brennan:2020ehu}.
The physical interpretation of the higher-group structures 
can be given
via
the Witten effect~\cite{Witten:1979ey} 
and the anomalous Hall effect for the axion.
Here, the Witten effect for the axion means that 
an axionic domain wall enclosing a magnetic monopole 
has an electric charge~\cite{Sikivie:1984yz}, 
and the anomalous Hall effect implies that there is an induced current 
when we add an electric field around an axionic string \cite{Qi:2008ew,Teo:2010zb,Wang:2012bgb}.
The purpose of this paper is to 
explore the higher-group structures of axion electrodynamics
by extending it to a generic $2n$-dimensional spacetime.

Higher-group structures can be efficiently described 
in the presence of the background gauge fields coupled with
the higher-form symmetry currents~\cite{Cordova:2018cvg}.
In \cite{Hidaka:2020izy}, the 3-group structure is realized as
a Green-Schwarz(GS)-type transformation \cite{Green:1984sg}
of the background gauge 
fields that are associated with Chern-Weil(CW) symmetry 
\cite{Heidenreich:2020pkc,Brauner:2020rtz}. 
The CW current is trivial
in that the current conservation is satisfied by Bianchi
identities, but plays a key role in understanding the 3-group structure.
In fact, the background gauging of the symmetries associated with 
equations of motion(EoMs) for the axion and photon
must require the simultaneous background gauging of CW symmetries with 
GS transformations enforced in order to preserve 
the gauge invariance of the axion and photon.
It is then natural to expect that higher-dimensional
axion electrodynamics exhibits enhanced algebraic structure compared
to four dimensions, because it admits a larger number of CW currents.

In this paper,
we study the higher-group structure of 
axion electrodynamics in $2n$ dimensions.
The higher group is organized by the CW symmetries and 
the higher-form symmetries associated to the equation of motion.
The 1-, 2-, $\cdots$, $(2n-2)$-form CW symmetries in
the $2n$-dimensional axion electrodynamics
must be gauged with the corresponding 
background gauge fields required to make a GS-type transformation 
in order to remove a quantum inconsistency due to
operator-valued ambiguities when we gauge the higher-form 
symmetries based on the equations of motion.
In particular, the $2n=6$ case is analyzed most intensively.
It is found that a new algebraic structure emerges such that
it contains the 3-group structure of the $2n=4$ case as a 
substructure with a trinary operation among three symmetry
generators encoded in it.

The organization of this paper is as follows.
Section \ref{2n} gives the action of the $2n$-dimensional axion
and Maxwell theory and then derives the global symmetries of the system.
It is found they are divided into two classes, one containing the symmetries
whose current is conserved by the equations of motion(EoMs) and
the other composed of the CW symmetry.
In section \ref{2nbkg}, 
we introduce the background gauge fields for the global
symmetries, and derive 
the GS-type transformation laws for the CW symmetry
gauge fields. They are the manifestation of higher-group structure
that contains the 3-group as a substructure.
Section \ref{n=3} focuses on the $2n=6$ case. Using the GS transformation
laws derived in section \ref{2nbkg}, we compute some correlation functions among the
symmetry generators of the 6-dimensional axion electrodynamics.
Section 5 is devoted to a conclusion and discussion.
In appendix A, we review a method developed in \cite{Hidaka:2020iaz}
for computing the correlation functions of the symmetry generators.

\section{$2n$-dimensional axion electrodynamics}
\label{2n}

The action of axion electrodynamics in $2n$ dimensions is given by
\begin{equation}
  S = - \int_{\cM_{2n}} \left( \frac{v^2}{2} |d\phi|^2 + \frac{1}{2e^2} |da|^2 -\frac{N}{(2\pi)^n n!}\phi\, (da)^n\right)\ .
  \label{action}
\end{equation}
Here, $\phi$ is an axion field with the $2\pi$ periodicity 
$\phi\sim\phi+2\pi$ 
and $a = a_\mu dx^\mu$ an $U(1)$ 1-form gauge field
with the Dirac quantization $\int_{\cal S} da \in 2\pi \bb{Z}$
for a closed 2-dimensional subspace ${\cal S}$. 
$N$ is quantized to be integer when $\cM_{2n}$ is a spin manifold.
We use the notation of differential forms.
The symbol $d$ denotes the exterior derivative, 
and $\wed $ is the wedge product.
The kinetic term of a $p$-form field ${\cal X}$
is written in terms of 
$|\cX|^2 = \cX \wedge \star \cX = 
 \fr{1}{p!} {\cal X}^{\mu_1 ...\mu_p}{\cal X}_{\mu_1 ...\mu_p} d^{2n} x$
with $\star$ being the Hodge star.
We sometimes 
abbreviate wedge products $\underbrace{da \wed \cdots \wed da}_{n} $ 
to $(da)^n$.
$v$ is a dimensionful parameter of mass dimension $n-1$ and $e$ the $U(1)$
gauge coupling constant. For simplicity, we set $v=1$ and $e=1$ hereafter.

\subsection{Symmetries}

In this subsection, we find out the global symmetries of
the action \eqref{action}.
As shown below, it possesses higher-form symmetries
of any integer ranks, which are divided into
two classes. One contains EoM-based discrete symmetries and the other
consists of CW symmetries.

\subsubsection*{EoM-based global symmetries}

The EoMs of $\phi$ and $a$ read
\begin{align}
  d\star d\phi=-\frac{N}{(2\pi)^nn!}(da)^n \ ,~~~
  d\star da=\frac{N}{(2\pi)^n(n-1)!}d\phi \wed (da)^{n-1} \ .
\end{align}
The EoM of $\phi$ leads to the 0-form symmetry current
\begin{equation}
  j_{\rm EoM}^{[0]} = -\star d\phi -\frac{N}{(2\pi)^n n!}a \wed 
(da)^{n-1} \ . 
\label{jpe}
\end{equation}
This defines the discrete 0-form symmetry $\bZ_N^{[0]}$.
To see this, we note that the symmetry generator, which is given by
exponentiating the current integrated over a $(2n-1)$-dimensional
manifold, is gauge invariant if the rotation angle is $\bZ_N$-valued:
\begin{equation}
  U^{[0]}_{\rm EoM}(\cD_{2n-1},\frac{2\pi m}{N}) = e^{ \frac{2i\pi m}{N} \int_{\cD_{2n-1}} j_{\rm EoM}^{[0]}} \ ,
\end{equation}
with $m\in \bZ$.
Here, $\cD_{2n-1}$ is a $(2n-1)$-dimensional closed subspace.

The EoM of the photon gives the current
\begin{equation}
    j_{\rm EoM}^{[1]} = \star da - \frac{N}{(2\pi)^n (n-1)!}\phi (da)^{n-1}\ . 
\label{jae}
\end{equation}
The Noether charge is obtained by integrating the current over
a $(2n-2)$-dimensional manifold $\cD_{2n-2}$ and generates the discrete
1-form symmetry $\bZ_{N}^{[1]}$
with the symmetry generator given by
\begin{equation}
  U^{[1]}_{\rm EoM}(\cD_{2n-2},\frac{2\pi m}{N}) = e^{ \frac{2i\pi m}{N} \int_{\cD_{2n-2}} j_{\rm EoM}^{[1]}}\ .
\end{equation}

\subsubsection*{Chern-Weil symmetries}

The Chern-Weil symmetry currents are conserved 
because of the Bianchi identities. The model (\ref{action})
has $2n$ CW currents, which generate
$(-1)$-, 0-, $\cdots$, $(2n-2)$-form $U(1)$ symmetries.
{}For an integer $r$ with $0\leq r \leq n-1$, 
the $2r$-form symmetry current is given by
\begin{equation}
  j_{\rm CW}^{[2r]} =\frac{1}{(2\pi)^{n-r}(n-r-1)!}\, d\phi \wed 
 (da)^{n-r-1}
\ . \label{jmeven}
\end{equation}
Also the $(2r-1)$-form symmetry current is given by
\begin{equation}
  j_{\rm CW}^{[2r-1]} =\frac{1}{(2\pi)^{n-r}(n-r)!}\,(da)^{n-r}\ . \label{jmodd}
\end{equation}
These yield the higher-form $U(1)$ symmetry generator of rank
$k=0,1,2,\cdots,2n-2$
\begin{equation}
  U_{\rm CW}^{[k]}(\cD_{2n-k-1},\gamma) = e^{i\gamma \int_{\cD_{2n-k-1}}j_{\rm CW}^{[k]}} \ ,
\end{equation}
with $\gamma\in\bR/2\pi\bZ$.

\subsection{Background gauging and 't Hooft anomaly}
\label{2nbkg}

Here, we gauge the higher-form symmetries by coupling the
currents found in the previous subsection to background gauge fields.
As found in \cite{Hidaka:2020izy}, gauging
$\bZ_{N}^{[0]}$ and $\bZ_{N}^{[1]}$
causes an operator-valued ambiguity, which 
states that the gauged action depends on how
it is extended to an extra dimension in order to
make it gauge invariant.
This problem is resolved by requiring that the background
gauge fields for the CW symmetries 
transform under
the $\bZ_{N}^{[0]}$ and $\bZ_{N}^{[1]}$ transformations
in an appropriate manner.

We first gauge $\bZ_{N}^{[1]}$, which acts on the photon field as
\begin{equation}
  a \rightarrow a + \Lambda_1 \ .
\end{equation}
with $\Lambda_1$ being a 1-form gauge transformation function.
We consider a background $\bZ_{N}$ gauge field 
given by a set of 2- and 1-forms $(B_2,B_1)$ with
$NB_2=dB_1$, 
which transform as
\begin{equation}
  B_2 \rightarrow B_2 + d\Lambda_1\ ,~~ \ B_1 \rightarrow B_1 + N\Lambda_1 \ .
\end{equation}
Here, $B_1$ is normalized as $\int_{\cal S} dB_1 \in 2\pi \bb{Z}$
for a 2-dimensional closed surface ${\cal S}$.
Then, $da$ in the action \eqref{action} should be replaced with $da - B_2$
in order to ensure that the action is gauge invariant.
The gauged action is in conflict with the periodicity of the axion field 
$\phi\rightarrow\phi+2\pi$, however. 
This problem can be rephrased as 
an ambiguity of how to extend the action in the partition function
defined on $\cM_{2n}$ to the action
defined on an artificial 
$(2n+1)$-dimensional manifold $\Omega_{\cM_{2n}}$ with 
$\partial\Omega_{\cM_{2n}}=\cM_{2n}$:
\begin{align}
  \frac{N}{(2\pi)^n n!}\int_{\cM_{2n}} \phi (da-B_2)^n 
= 
\frac{N}{(2\pi)^n n!}\int_{\Omega_{\cM_{2n}}} d\phi \wed (da-B_2)^n
\quad\text{mod }2\pi.
  \label{topo2n+1}
\end{align}
Hereafter, we omit writing ``mod $2\pi$'' for simplicity.
{}For the purpose of computing the difference of the actions that arises
from two choices of $(2n+1)$-manifolds $\Omega_{\cM_{2n}}$ and 
$\Omega'_{\cM_{2n}}$, we define a closed manifold 
$\cZ_{2n+1}=\Omega_{\cM_{2n}}\sqcup \overline{\Omega'_{\cM_{2n}}}$
and evaluate the gauged topological action on it.
Here, $\overline{\Omega'_{\cM_{2n}}}$ is $\Omega'_{\cM_{2n}}$
with an opposite orientation.
By expanding it as
\begin{align}
  \frac{N}{(2\pi)^n n!}\int_{\Omega_{\cM_{2n}}} d\phi\wed (da-B_2)^n
=\frac{N}{(2\pi)^n}\sum_{k=0}^n
\frac{(-1)^{n-k}}{k!(n-k)!}\int_{\Omega_{\cM_{2n}}} d\phi\wed (da)^k 
\wed (B_2)^{n-k}
\end{align}
only the 0th and 1st order terms in $B_2$ take values in $2\pi\bZ$ 
because of the normalization condition
$N\int B_2\in 2\pi\bZ$.
The operator-valued ambiguities originate from the rest of the terms that are
nonlinear in $B_2$.

We now show that the quantum ambiguities can be eliminated by
adding the local counterterms
\begin{align}
& \sum_{r=1}^{n-1}\int_{\Omega_{\cM_{2n}}} j_{{\rm CW}}^{[2r]}\big|_{da\to da-B_2}\wedge Y_{2r+2}
\nn\\
  =& \sum_{k=0}^{n-2}\int_{\Omega_{\cM_{2n}}}
 d\phi\, (da)^k \sum_{r=1}^{n-k-1}\frac{(-1)^{n-k-r}}{(2\pi)^{n-r}k!(n-k-r-1)!}(B_2)^{n-r-k-1}\wedge Y_{2r+2} \ .
 \label{counter1}
\end{align}
The replacement $da\to da-B_2$ should be made in order to keep the counterterm
term invariant under the $\bZ_{N}^{[1]}$ gauge 
transformation.
It is interesting to note that
no replacement is necessary for the four-dimensional axion-Maxwell system because
the quantum ambiguities for the $2n=4$ case that involve the axion field
are independent of $da$. This is manifest also upon setting $n=2$ in 
\eqref{counter1}.
$Y_{2r+2}$ is a $(2r+2)$-form field strength of the form 
$Y_{2r+2}=dX_{2r+1} + \alpha_{2r+2}$.
$X_{2r+1}$ is the background CW gauge field that couples minimally with
the current $j_{\rm CW}^{[2r]}$
with the normalization condition given by
$\int dX_{2r+1}\in 2\pi\bZ$. $\alpha_{2r+2}$ is fixed
by requiring that 
the local counter terms cancel the operator-valued ambiguities from
the gauged topological term.
To see this, we note that
adding \eqref{counter1} to the topological term \eqref{topo2n+1} gives
the integrand
\begin{align}
& \sum_{k=0}^{n-2}
\frac{1}{(2\pi)^{k+1}\,k!}\,d\phi\,(da)^k
\nn\\
&
\qquad\wedge\left[
2\pi N\,\frac{(-1)^{n-k}}{(2\pi)^{n-k}\,(n-k)!}
(B_2)^{n-k}\right.
\nn 
\\
&
\left.
\hphantom{\qquad\wedge\qquad}
-\sum_{r=1}^{n-k-1}\frac{(-1)^{n-k-r-1}}
{(2\pi)^{n-k-r-1}\,(n-k-r-1)!}\,(B_2)^{n-k-r-1}\wedge Y_{2r+2}
\right]
\nonumber\\
+&
\frac{N}{(2\pi)^n}\sum_{k=n-1}^n
\frac{(-1)^{n-k}}{k!(n-k)!} d\phi\, (da)^k\,(B_2)^{n-k}.
  \label{OVA+Counter_1-form}
\end{align}
The operator-valued ambiguities can be cancelled
by requiring
\begin{align}
&
2\pi N\,\frac{(-1)^{n-k}}{(2\pi)^{n-k}\,(n-k)!}
(B_2)^{n-k}
-\sum_{r=1}^{n-k-1}\frac{(-1)^{n-k-r-1}}
{(2\pi)^{n-k-r-1}\,(n-k-r-1)!}\,(B_2)^{n-k-r-1}\wedge Y_{2r+2}
\nn
\\
&=-dX_{2(n-k)-1} \ ,  
\label{alpha2r+2}
\end{align}
for $k=0,1,2,\cdots,n-2$.
This allows us to fix $\alpha_{2r+2}$ completely.
We start examining the cancellation condition
from the $k=n-2$ case to obtain
$\alpha_4 = \frac{N}{4\pi}(B_2)^2$.
The rest of $\alpha_{2r+2}$ can be worked out by solving
the cancellation conditions for $k=n-3,n-4,\cdots,0$ recursively.
As clear from the sample computation of $\alpha_4$, 
all $\alpha_{2r+2}$ are written in terms of
$B_2$. Naively, this implies that $Y_{2r+2}$ is not $\bZ_{N}^{[1]}$ 
gauge invariant so that the local counter terms break
the $\bZ_{N}^{[1]}$ gauge invariance.
It is restored by requiring that $X_{2r+1}$ transform under
$\bZ_{N}^{[1]}$ in such a manner that $Y_{2r+2}$ is kept
gauge invariant. This is the reason for why the GS-type transformation
must be imposed for $X_{2r+1}$.

{}Furthermore, we gauge $\bZ^{[0]}_N$, which acts on the axion as a shift:
\begin{equation}
  \phi \rightarrow \phi + \Lambda_0 \ .
\end{equation}
The $\bZ^{[0]}_N$ background gauge field is defined by a pair
of 1- and 0-form fields 
$(A_1,A_0)$ with $NA_1=dA_0$, and the $\bZ^{[0]}_N$ gauge transformation
acts as
\begin{equation}
  A_1 \rightarrow A_1 + d\Lambda_0\ ,~~ \ A_0\rightarrow A_0 + N\Lambda_0
\ .
\end{equation}
Here, $A_0$ is normalized as $\int_{\cal C} dA_0 \in 2\pi \bb{Z}$ 
for a closed 1-dimensional subspace ${\cal C}$.
The $\bZ^{[0]}_N$ gauge invariant action is obtained by
replacing $d\phi$ with the covariant derivative
$d\phi - A_1$.
Then, 
\eqref{OVA+Counter_1-form} together with \eqref{alpha2r+2}
leads to the linear term in $A_1$:
\begin{align}
&  \frac{1}{2\pi}\,A_1 \wed dX_{2n-1}
+\sum_{k=1}^{n-2}\frac{1}{(2\pi)^{k+1}\,k!}
\,A_1\wed (da)^k\wed dX_{2(n-k)-1}
\nn
\\
&\quad
+\frac{N}{(2\pi)^n\,(n-1)!}\,A_1\wed (da)^{n-1}\wed B_2
-\frac{N}{(2\pi)^n\,n!}\,A_1\wed (da)^{n} \ .
\label{linearA1}
\end{align}
It is clear that the second and the third terms are responsible for
another quantum ambiguity. The first term is written only by
the background field, resulting in an 't Hooft anomaly.
This ambiguity can be canceled by further adding the counterterms
\begin{align}
&\sum_{r=1}^{n-1} j_{{\rm CW}}^{[2r-1]}\big|_{da\to da-B_2}\wedge Y_{2r+1} 
\nonumber\\
  =& 
\sum_{r=1}^{n-1}\frac{(-1)^{n-r}}{(2\pi)^{n-r}\,(n-r)!}(B_2)^{n-r}\wedge
Y_{2r+1}
\nn
\\
&
+
\sum_{k=1}^{n-1}\frac{1}{(2\pi)^k\,k!}(da)^k
 \sum_{r=1}^{n-k} \frac{(-1)^{n-k-r}}{(2\pi)^{n-k-r}\,(n-k-r)!}(B_2)^{n-k-r}\wedge Y_{2r+1} \ ,
  \label{counter2}
\end{align}
where $Y_{2r+1}=dX_{2r}+\alpha_{2r+1}$ with the normalization condition
given by $\int dX_{2r+1}\in 2\pi\bZ$.
Adding it to (\ref{linearA1}) cancels the ambiguity by requiring that
$\alpha_{2r+1}$ obey
\begin{align}
dX_2&=\frac{N}{2\pi}\,A_1 \wed B_2+Y_3 \ ,
\label{X2}
\\
dX_{2(n-k)}&=\frac{1}{2\pi}\,A_1 \wed dX_{2(n-k)-1}
+\sum_{r=1}^{n-k}\frac{(-1)^{n-k-r}}{(2\pi)^{n-k-r}\,(n-k-r)!}
\,(B_2)^{n-k-r}\wedge Y_{2r+1} \ ,
\label{X2(n-k)}
\end{align}
for $k=1,2,\cdots,n-2$.
(\ref{X2}) determines $\alpha_3$ and (\ref{X2(n-k)}) can be solved
recursively as before to fix the rest of $\alpha_{2r+1}$.
{}For instance, the condition for $k=n-2$ is solved by setting
\begin{align}
  \alpha_5=-\frac{1}{2\pi}\,A_1\wed dX_3
+\frac{1}{2\pi}\,B_2 \wed \left(dX_2-\frac{N}{2\pi}\,A_1 \wed B_2\right)
\ .
\end{align}
It is found that $\alpha_{2r+1}$ is not gauge invariant either.
The counterterm \eqref{counter2} is left gauge invariant
by requiring that $X_{2r}$ make a GS-type transformation 
under $\bZ_{N}^{[0]}$ and $\bZ_{N}^{[1]}$
in such a manner that $Y_{2r+1}$ becomes a gauge invariant field 
strength.

To summarize, the gauge invariant action with no
operator-valued ambiguity is given by
\begin{align}
  S^{\prime} =&- \int_{\cM_{2n}} \left( \frac{1}{2} |d\phi-A_1|^2 + \frac{1}{2} |da-B_2|^2 \right) \nonumber \\
  &+ \frac{N}{(2\pi)^n n!} \int_{\Omega_{\cM_{2n}}} (d\phi-A_1) \wedge (da-B_2)^n + 
\sum_{r=1}^{2n-2}\int_{\Omega_{\cM_{2n}}} 
j_{\rm CW}^{[r]}\Big|_{d\phi\to d\phi-A_1,\,da\to da-B_2}\wedge Y_{r+2}
\ .
  \label{gauged}
\end{align}
This might depend on the choice of ${\Omega_{\cM_{2n}}}$, which
is used to rewrite Chern-Simons(CS) terms in a 
gauge invariant manner. The difference of the actions
for two choices of $\Omega_{\cM_{2n}}$ 
is manifested as a 't Hooft anomaly.
More concretely, we define the compact $(2n+1)$-dimensional
manifold $\cZ_{2n+1}$ such that 
$\cZ_{2n+1}=\Omega_{\cM_{2n}}^{(1)} \cup\overline{\Omega_{\cM_{2n}}^{(2)}}$,
where $\Omega_{\cM_{2n}}^{(1)}$ and $\Omega_{\cM_{2n}}^{(2)}$ 
are glued together at the common boundary 
$\partial\Omega_{\cM_{2n}}^{(1)}=\partial\Omega_{\cM_{2n}}^{(2)}=\cM_{2n}$. 
Then, the 't Hooft anomaly is given by a phase
in the partition function, 
\begin{align}
  \exp i\int_{\cZ_{2n+1}}\left[
  \frac{1}{2\pi}A_1 \wed dX_{2n-1}
+\sum_{r=1}^{n-1}\frac{(-1)^{n-r}}{(2\pi)^{n-r}\,(n-r)!}\,(B_2)^{n-r}
\wed Y_{2r+1} 
\right] \ .
\label{thooft:n}
\end{align}
Here, the integrand comes from the terms 
in (\ref{gauged}) that depend only on the background gauge
field.
If $\cZ_{2n+1}$ is taken to be a mapping torus that interpolates
between two $\cM_{2n}$, each of which is endowed with
background gauge fields related to each other via
a gauge transformation,
the phase \eqref{thooft:n} leads to an anomalous phase
associated with the gauge transformation.

The explicit form of the 't Hooft anomaly for the $n=3$ case
is given in the next section.

%%%%%%%%%%%%%%%%%%%%%%%%%%%%%%%%%%%%%%%%%%%%%%%%%%%%%%%%%%%%%%%%%%%%%%%%%%%%%%%%%%
\section{The $n=3$ case}
\label{n=3}

In this section, we make a detailed analysis of axion 
electrodynamics in six dimensions.
This is in parallel with that made
in \cite{Hidaka:2020iaz,Hidaka:2020izy} for the $2n=4$ case.

The action in the absence of the background gauge field
reads
\begin{equation}
  S_{\rm 6d}=-\int_{\cM_6} \left( \frac{1}{2}\left|d\phi \right|^2 + \frac{1}{2}\left|da \right|^2 -\frac{N}{48\pi^3} \phi \left(da\right)^3 \right)
\ .
  \label{6dimaction}
\end{equation}
Here, we briefly discuss the Witten effect induced on an axionic 
domain wall and the anomalous Hall effect by an axionic
vortex. 
A more rigorous analysis will be given later.
The EoMs for $\phi$ and $a$ are given in component by
\begin{align}
  \partial^2 \phi &= -\frac{N}{384 \pi^2}\epsilon^{\mu\nu\rho\sigma\alpha\beta} F_{\mu\nu}F_{\rho\sigma}F_{\alpha\beta}\ ,
\label{eom:phi}\\
  \partial_\mu F^{\mu\nu} &= \frac{N}{64 \pi^2}\epsilon^{\nu\rho\sigma\tau\alpha\beta} (\partial_\rho\phi)F_{\sigma\tau}F_{\alpha\beta}
\ .\label{eom:a}
\end{align}
Assuming that $F_{\mu\nu}$ is static, the EoM \eqref{eom:a} for $\nu=0$ 
gives
\begin{align}
\partial_i F^{i0} = \frac{N}{64 \pi^2}\epsilon^{0ijklm} (\partial_i\phi)F_{jk}F_{lm}
\ ,\nn
\end{align}
with $i,j,k,\cdots$ being the indices for the spatial directions.
By turning on a domain wall configuration for $\phi$ together with
a magnetic flux over the spatial direction of the domain wall,
the RHS serves as an electric charge density induced on
the domain wall. This is the Witten effect.

The EoM \eqref{eom:a} for $\nu=m$ becomes
\begin{align}
\partial_n F^{nm} = -\frac{N}{64 \pi^2}\epsilon^{0ijklm}(\partial_i \phi)
F_{0j} F_{kl}
\ .\nn  
\end{align}
We give a vortex configuration to $\phi$, which appears as an
axionic 3-brane. In addition, we turn on an electric 
field and a magnetic flux on the 3-brane world volume so that
the RHS is nonvanishing.
Then, the resultant source term serves as an electric current
that flows towards the 3-brane, which is normal to
the electric field direction.

The symmetries and the corresponding currents coupled minimally with
the background gauge fields are listed as
\begin{table}[H]
  \centering
  \begin{tabular}{|c|c|l|c|} 
  \hline
      Generator & Group & Current & Gauge field\\
  \hline
      $U_{\rm EoM}^{[0]}$ & $\mathbb{Z}_N$ & $j_{\rm EoM}^{[0]} = -\star d\phi - \frac{N}{48\pi^3} a\wedge da\wedge da$ & $A_1$\\
  \hline
      $U_{\rm EoM}^{[1]}$ & $\mathbb{Z}_N$ & $j_{\rm EoM}^{[1]} = \star da-\frac{N}{16\pi^3} \phi\wedge da\wedge da$ & $B_2$\\
  \hline
      $U_{\rm CW}^{[1]}$ & $U(1)$ & $j_{\rm CW}^{[1]} = \frac{1}{8\pi^2}da\wedge da$ & $B_2^{\rm CW}$\\
  \hline
      $U_{\rm CW}^{[2]}$ & $U(1)$ & $j_{\rm CW}^{[2]} = \frac{1}{4\pi^2}d\phi\wedge da$ & $C_3$\\
  \hline
      $U_{\rm CW}^{[3]}$ & $U(1)$ & $j_{\rm CW}^{[3]} = \frac{1}{2\pi}da$ & $D_4$\\
  \hline
      $U_{\rm CW}^{[4]}$ & $U(1)$ & $j_{\rm CW}^{[4]} = \frac{1}{2\pi}d\phi$ & $E_5$\\
  \hline
  \end{tabular}
\end{table}
Here, the gauge invariant field strengths $Y_3,Y_4,Y_5,Y_6$
are renamed as $G_3,H_4,I_5,J_6$ respectively:
\begin{align}
  G_3 &= dB_2^{\rm CW} - \frac{N}{2\pi} A_1 \wedge B_2, \label{G3}\\
  H_4 &= dC_3 + \frac{N}{4\pi} B_2 \wedge B_2, \label{H4}\\
  I_5 &= dD_4 - \frac{1}{2\pi} A_1 \wedge dC_3 + \frac{1}{2\pi} B_2 \wedge dB_2^{\rm CW} -\frac{N}{4\pi^2} A_1 \wedge B_2 \wedge B_2,\label{I5}\\
  J_6 &= dE_5 + \frac{1}{2\pi} B_2\wedge dC_3 + \frac{N}{12\pi^2} B_2 \wedge B_2 \wedge B_2,\label{J6}
\end{align}
As seen before, the gauge fields make a GS transformation in such a manner
that these are left invariant under the $\bZ_N^{[0]}$ and
$\bZ_N^{[1]}$ gauge transformations:
\begin{align}
  B_2^{\rm CW} \rightarrow & B_2^{\rm CW} + d\Lambda_1^{\rm CW} - \frac{N}{2\pi}\left(A_1 + d\Lambda_0\right)\wedge \Lambda_1 + \frac{N}{2\pi}B_2 \wedge \Lambda_0\ , \label{B}\\
  C_3 \rightarrow & C_3 + d\Lambda_2 -\frac{N}{4\pi} \left(B_2 + d\Lambda_1\right)\wedge \Lambda_1 -\frac{N}{4\pi} B_2 \wedge \Lambda_1\ , \label{C}\\
  D_4 \rightarrow & D_4 + d\Lambda_3 - \frac{1}{2\pi}d\Lambda_0\wedge C_3 -\frac{1}{2\pi}d\Lambda_1 \wedge B_2^{\rm CW} \nonumber\\
  & -\frac{N}{8\pi^2}\left(d\Lambda_1 \wedge (B_2+d\Lambda_1)\wedge d\Lambda_0 + d\Lambda_1\wedge A_1 \wedge d\Lambda_1 +d\Lambda_1\wedge B_2 \wedge d\Lambda_0\right)\ , \label{D}\\
  E_5 \rightarrow & E_5 + d\Lambda_4 - \frac{N}{8\pi^2} B_2 \wedge \left(B_2 + d\Lambda_1\right) \wedge \Lambda_1 - \frac{N}{12\pi^2} d\Lambda_1 \wedge d\Lambda_1 \wedge \Lambda_1\nonumber\\
  & -\frac{1}{2\pi}d\Lambda_1 \wedge C_3 +\frac{N}{8\pi^2}(B_2+d\Lambda_1)^2 \wedge \Lambda_1\ . \label{E}
\end{align}

The partition function of the 6d axion-Maxwell system, which 
is a functional
of the background gauge fields,
is given by
\begin{equation}
  Z[A_1, B_2, B_2^{\rm CW}, C_3, D_4, E_5] = \int\cD\phi\,\cD a\,\exp
i\left[S_{\rm 6d} +S_{\rm min} +\int_{\Omega_{\cM_6}} \cL_{7}\right]
\ ,
  \label{pf}
\end{equation}
with
\begin{equation}
  S_{\rm min}=\int_{\cM_6} \left(j_{\rm EoM}^{[0]}\wedge A_1 + j_{\rm EoM}^{[1]}\wedge B_2 + j_{\rm CW}^{[1]}\wedge B_2^{\rm CW} + j_{\rm CW}^{[2]}\wedge C_3+ j_{\rm CW}^{[3]}\wedge D_4+ j_{\rm CW}^{[4]}\wedge E_5 \right)\ ,
\end{equation}
\begin{equation}
  \cL_{7}=\frac{N}{16\pi^3} A_1\wedge (B_2)^3 -\frac{1}{8\pi^2} (B_2)^2 \wedge dB_2^{\rm CW} +\frac{1}{4\pi^2} A_1 \wedge B_2 \wedge dC_3 -\frac{1}{2\pi} B_2 \wedge dD_4 +\frac{1}{2\pi}A_1\wedge dE_5
\ .
  \label{anomaly6d}
\end{equation}
Here, the 't Hooft anomaly for $n=3$ is determined by the difference
of \eqref{anomaly6d} due to two choices of $\Omega_{\cM_6}$,
see \eqref{thooft:n}. The seagull terms
proportional to $|B_2|^2$ and $|A_1|^2$ are omitted for simplicity
because these are irrelevant to the rest of the discussions in
this paper.

\subsection{Charged objects}

Here, we discuss what are objects charged
under the global symmetries we found before, and
then compute the charges explicitly.
In \cite{Hidaka:2020izy,Hidaka:2020iaz}, these are
obtained by computing correlation functions
involving charged objects and symmetry generators. 
This method
is reviewed in the appendix \ref{alt}.
In this paper, we give an alternative prescription
based on a systematic use of the background gauge
fields.
It is found that
the global charges are
worked out from the 't Hooft anomaly.

\subsubsection{EoM-based symmetries}

We first identify charged objects under
the EoM-based global symmetries $\bZ_{N}^{[0]}$ and $\bZ_{N}^{[1]}$.

$\bZ_{N}^{[0]}$ acts on $\phi$ as a shift so that
the associated charged object is given by a local operator
$\sfI(\cP_{\rm defect},q)\equiv e^{iq\phi(\cP_{\rm defect})}$.
Here, $\cP_{\rm defect}$ is a point in the spacetime 
${\cal M}_6$ on which the local 
operator is localized with $q\in \bZ$ being the $\bZ_{N}^{[0]}$ charge.
Using 
\begin{align}
e^{iq\phi(\cP_{\rm defect})}=e^{2\pi i q\int j_{\rm CW}^{[4]} \wedge \delta_5(\Omega_{\cP_{\rm defect}})}\ ,\nn
\end{align}
the local operator arises by turning on 
$E_5=2\pi q \delta_5(\Omega_{\cP})$.
Here, the delta function $\delta_p(\cD_{D-p})$ with the support $\cD_{D-p}$ 
being a submanifold of codimension $p$ is defined to satisfy 
\begin{align}
\int_{\cM_D} J^{(D-p)}\wedge \delta_p(\cD_{D-p})\equiv \int_{\cD_{D-p}} J^{(D-p)}\ ,\nn
\end{align}
for any $(D-p)$-form $J^{(D-p)}$.
Use of the Stokes theorem yields
$\delta_{p}({\cD}_{D-p})= (-1)^{D-p-1} d\delta_{p-1} (\Omega_{{\cD}_{D-p}})$
with $\partial\Omega_{{\cD}_{D-p}}={\cD}_{D-p}$.

We now argue that the 0-form symmetry generator 
$U_{\rm EoM}^{[0]}(\cV,2\pi n/N)$,  which is supported on a codimension-one
manifold $\cV$, acts on $\sfI(\cP_{\rm defect},q)$ nontrivially.
This is done by defining the correlation
functions 
$\left\langle U_{\rm EoM}^{[0]}\left(\cV,\frac{2\pi n}{N}\right) \sfI\left(\cP_{\rm defect},q\right) \right\rangle$
and
$\left\langle \sfI\left(\cP_{\rm defect},q\right) \right\rangle$
from the partition function \eqref{pf} in the presence of
the appropriate background gauge fields and then
relating them with each other 
via the 't Hooft anomaly.
We define the correlator 
$\left\langle U_{\rm EoM}^{[0]}\left(\cV,\frac{2\pi n}{N}\right) \sfI\left(\cP_{\rm defect},q\right) \right\rangle$
by inserting 
$A_1=\frac{2\pi n}{N}\delta_1(\cV)$ and 
$E_5=2\pi q \delta_5(\Omega_{\cP_{\rm defect}})$
into \eqref{pf}:
\begin{equation}
\left\langle U_{\rm EoM}^{[0]}\left(\cV,\frac{2\pi n}{N}\right) \sfI\left(\cP_{\rm defect},q\right) \right\rangle\
\equiv Z\left[\frac{2\pi n}{N}\delta_1(\cV),0,0,0,0,2\pi q \delta_5(\Omega_{\cP_{\rm defect}})\right] \ .
\end{equation}
Note that 
$A_1=\frac{2\pi n}{N}\delta_1(\cV)$
leads to a codimension-1 defect, which is referred to
as an axionic domain wall.
The relation between the axionic domain wall
and $U_{\rm EoM}^{[0]}\left(\cV,\frac{2\pi n}{N}\right)$
will be important later.
We next make a gauge transformation to gauge away $A_1$, which
amounts to eliminating $U_{\rm EoM}^{[0]}\left(\cV,\frac{2\pi n}{N}\right)$.
It follows from \eqref{B}-\eqref{E} that
this induces no CW gauge fields.
Then, there remains only $E_5$, in the presence of which
the partition function defines 
$\left\langle \sfI\left(\cP_{\rm defect},q\right) \right\rangle$:
\begin{align}
\left\langle \sfI\left(\cP_{\rm defect},q\right) \right\rangle
\equiv  Z[0,0,0,0,0,2\pi q \delta_5(\Omega_{\cP_{\rm defect}})] \ .
\end{align}
{}Finally, we evaluate the 't Hooft anomaly associated with the
gauge transformation. As discussed in \eqref{thooft:n} and \eqref{anomaly6d},
it is obtained by integrating $\cL_{7}$ over a mapping 7-torus
$\cT_7$ of topology $\cM_6\times\bS^1$,
where $\bb{S}^n$ denotes a $n$-dimensional sphere.
Let $\cA_1$ and $\cE_5$ be an lift of the background gauge field
$A_1$ and $E_5$ respectively to $\cT_7$ such that
\begin{equation}
  \cA_1 = (1-\tau)A_1\ ,~~~\cE_5 = E_5 \ ,
\end{equation}
with $\tau\in[0,1]$ being the coordinate of $\bS^1$.
It is easy to find that
\begin{align}
\exp\left( i\int_{\cT_7} \cL_{7} \right)
=
  \exp\left( \frac{i}{2\pi} \int_{\cT_7} \cA_1 \wedge d\cE_5 \right)
= \exp\left({\frac{2\pi iqn}{N}\int_{\cM_6}\delta_1(\cV) \wedge \delta_5(\Omega_{\cP_{\rm defect}})}\right)
\ .
\end{align}
The resultant phase factor measures
the linking number between $\cV$ and $\cP_{\rm defect}$.
We thus find that
\begin{equation}
  \left\langle U_{\rm EoM}^{[0]}\left(\cV,\frac{2\pi n}{N}\right) \sfI\left(\cP_{\rm defect},q\right) \right\rangle =e^{\frac{2\pi iqn}{N}{\rm Link}\left(V,\cP_{\rm defect}\right)} \left\langle \sfI\left(\cP_{\rm defect},q\right) \right\rangle \ .
\end{equation}

The EoM-based 1-form symmetry $\bZ_N^{[1]}$ acts on $a$ as a shift and
gives rise to a nontrivial transformation for
the Wilson loop $\sfW(\cL_{\rm defect},q)=e^{iq\int_{\cL_{\rm defect}} a}$.
By noting
\begin{align}
  e^{iq\int_{\cL_{\rm defect}} a}=e^{2\pi i q\int j_{\rm CW}^{[3]} \wedge \delta_4(\Omega_{\cL_{\rm defect}})} \ ,
\end{align}
the Wilson loop operator is realized by turning on
$D_4 = 2\pi q\delta_4(\Omega_{\cL_{\rm defect}})$.
The partition function with
$B_2=\frac{2\pi m}{N}\delta_1(\cW)$ and 
$D_4 = 2\pi q\delta_4(\Omega_{\cL_{\rm defect}})$
defines the correlator 
$\left\langle U_{\rm EoM}^{[1]}\left(\cW,\frac{2\pi m}{N}\right) \sfW\left(\cL_{\rm defect},q\right) \right\rangle$.
By gauging away $B_2$ and evaluating the associated 't Hooft anomaly,
we obtain
\begin{equation}
  \left\langle U_{\rm EoM}^{[1]}\left(\cW,\frac{2\pi m}{N}\right) \sfW\left(\cL_{\rm defect},q\right) \right\rangle =e^{\frac{2\pi iqm}{N}{\rm Link}\left(\cW,\cL_{\rm defect}\right)} \left\langle \sfW\left(\cL_{\rm defect},q\right) \right\rangle \ .
\end{equation}

\subsubsection{Chern-Weil symmetries}\label{CWCO}

Here, we argue that
the charged objects for the 
Chern-Weil symmetries are composed of
axionic vortices and monopoles.

We first consider 
the 4-form symmetry generator 
$U^{[4]}_{\rm CW}(\cL,\alpha)$ with $\cL$ being a 1-dimensional support.
As the corresponding CW current is given by $d\phi/(2\pi)$,
the charged operator is realized by an axionic vortex, which
is equivalent to turning on $A_1$ such that
\begin{equation}
  \oint_{\bS^1}A_{1,{\rm defect}} = 2\pi q
\ .
\end{equation}
This defines the surface operator
$\sfV(\cW_{\rm defect},q)$
with $\cW_{\rm defect}$ being a codimension-2 support.
This is also interpreted as an axionic 3-brane.
\begin{figure}[H]
  \centering
  \includegraphics[width=5cm]{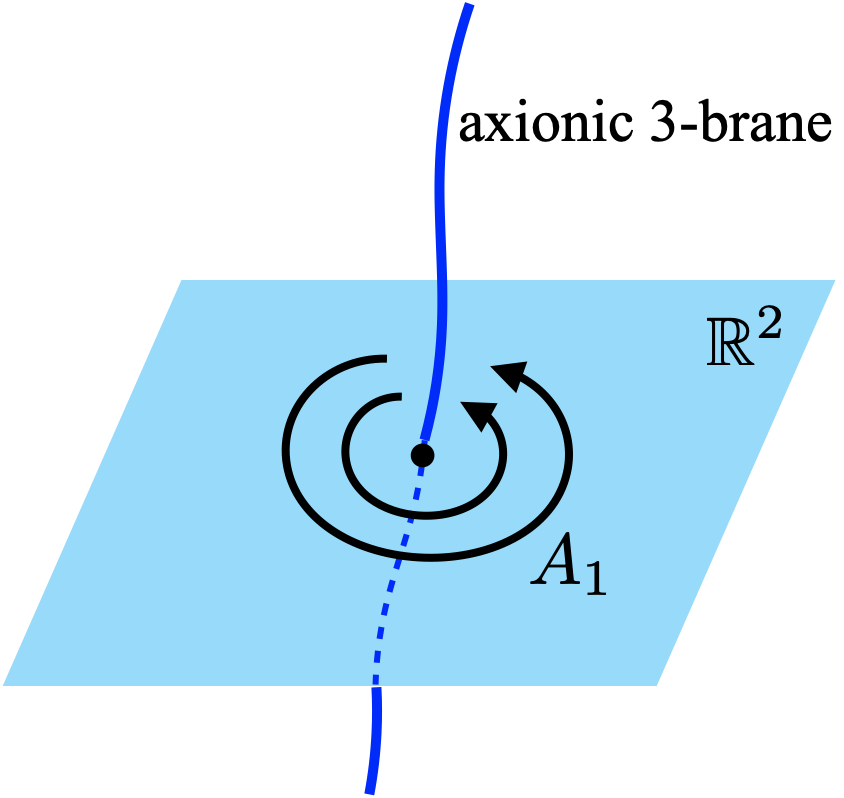}
  \caption{$\sfV(\cW_{\rm defect},q)$ as an axionic 3-brane
carrying the $U^{[4]}_{\rm CW}$ charge}
  \label{4form}
\end{figure}
We verify that the $U_{\rm CW}^{[4]}$ charge 
of $\sfV(\cW_{\rm defect},q)$ is computed by evaluating
the 't Hooft anomaly in \er{anomaly6d}.
The correlator
$\left\langle U^{[4]}_{\rm CW}\left(\cL,\alpha\right) \sfV(\cW_{\rm defect},q) \right\rangle$
is defined from the partition function with
$A=A_{1,{\rm defect}}$ and $E_5 =\alpha\delta_5(\cL)$:
\begin{equation}
  Z[A_{1,{\rm defect}},0,0,0,0,\alpha\delta_5(\cL)] =  \left\langle U^{[4]}_{\rm CW}\left(\cL,\alpha\right) \sfV(\cW_{\rm defect},q) \right\rangle \ .
\end{equation}
By making a gauge transformation to turn off $A_{1,{\rm defect}}$, we find 
\begin{equation}
  Z[A_{1,{\rm defect}},0,0,0,0,\alpha\delta_5(\cL)]   
=e^{i\alpha q {\rm Link}(\cW_{\rm defect},\cL)}
\, Z[0,0,0,0,0,\alpha\delta_5(\cL)] \ .
\end{equation}
Here, the phase factor in the RHS follows from the
't Hooft anomaly, which is computed by constructing a mapping
7-torus associated with the gauge transformation under consideration.
As the partition function in the RHS gives the one-point 
function of $\sfV(\cW_{\rm defect},q)$, we obtain
\begin{equation}
    \langle U^{[4]}_{\rm CW}\left(\cL,\alpha\right) \sfV(\cW_{\rm defect},q) \rangle 
=e^{i\alpha q {\rm Link}(\cW_{\rm defect},\cL)}\,
 \langle \sfV(\cW_{\rm defect},q) \rangle\ .
\end{equation}

Next, we discuss the 3-form symmetry generator 
$U^{[3]}_{\rm CW}(\cS,\beta)$, which is supported on a
codimension-4 manifold $\cS$.
As the corresponding CW current is $da/(2\pi)$, 
the charged object is a monopole.
This is realized by turning on the background gauge field
\begin{equation}
  \oint_{\bS^2}B_{2,{\rm defect}} = 2\pi q \ ,
\end{equation}
and defines a codimension-3 surface operator
$\sfT(\cC_{\rm defect},q)$ with $\cC_{\rm defect}$ being the worldvolume.
We call it an 't Hooft surface.
\begin{figure}[H]
  \centering
  \includegraphics[width=5cm]{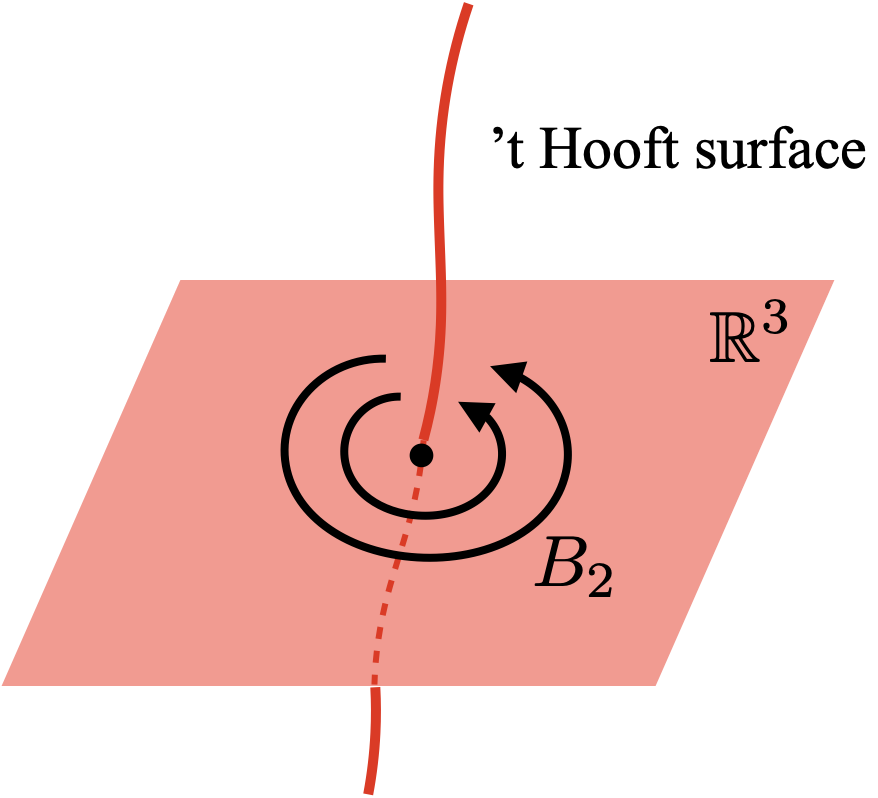}
  \caption{'t Hooft surface operator $\sfT(\cC_{\rm defect},q)$
carrying the $U^{[3]}_{\rm CW}$ charge}
  \label{3form}
\end{figure}
The $U(1)^{[3]}$ charge of the 't Hooft surface 
is computed from the 't Hooft anomaly.
{}For this purpose, we define the correlator of
$U^{[3]}_{\rm CW}(\cS,\beta)$ and
$\sfT(\cC_{\rm defect},q)$ as the partition function
in the presence of $B_{2,{\rm defect}}$ and
$D_4=\beta\delta_4(\cS)$:
\begin{align}
Z[0,B_{2,{\rm defect}},0,0,D_4=\beta\delta_4(\cS),0] 
= \langle U^{[3]}_{\rm CW}\left(\cS,\beta\right) \sfT(\cC_{\rm defect},q) 
\rangle  \ .
\end{align}
By gauging away $D_4$ and evaluating the associated 't Hooft anomaly,
we obtain
\begin{equation}
  \langle U^{[3]}_{\rm CW}\left(\cS,\beta\right) \sfT(\cC_{\rm defect},q) \rangle 
= e^{i\beta q {\rm Link}(\cC_{\rm defect},\cS)}
Z[0,B_{2,{\rm defect}},0,0,0,0] 
= e^{i\beta q {\rm Link}(\cC_{\rm defect},\cS)}\langle \sfT(\cC_{\rm defect},q) \rangle\ .
\end{equation}

The charged operator for the 
2-form symmetry $U(1)^{[2]}$
has a 2-dimensional worldvolume and is composed of
an 't Hooft surface and an axionic 3-brane wrapped around it,
because the corresponding CW current is given by $d\phi\wedge da/(2\pi)^2$.
\begin{figure}[H]
  \centering
  \includegraphics[width=5cm]{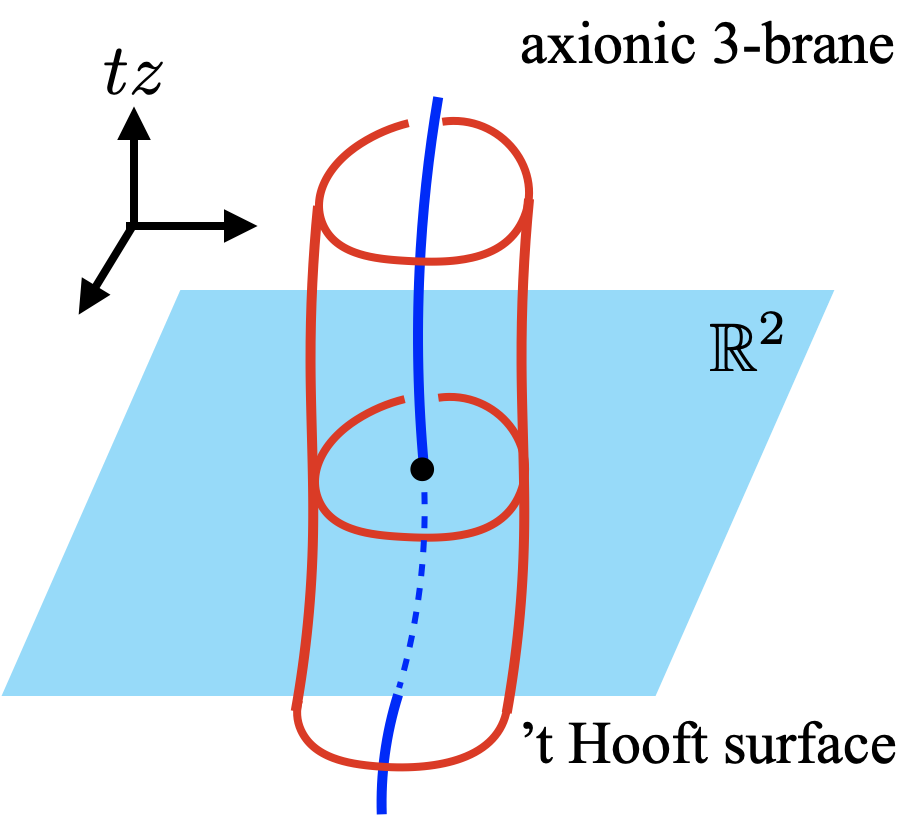}
  \caption{$\sfV\sfT(\cS_{\rm defect},q)$ 
carrying the $U^{[2]}_{\rm CW}$ charge}
  \label{2form}
\end{figure}
A typical configuration of the charged operator is listed below.
Here, $r$ and $\vartheta$ are the polar coordinates of the 2-dimensional
plane transverse to the axionic 3-brane.
\begin{table}[H]
  \centering
  \begin{tabular}{|c||c|c|c|c|c|c|} 
  \hline
       & $t$ & $x$ & $y$ & $z$ & $r$ & $\vartheta$\\
  \hline
      $\sfV\left(\cW_{\rm defect},q_\phi\right)$& $\circ$ & $\circ$ & $\circ$ & $\circ$ &  & \\
  \hline
      $\sfT\left(\cC_{\rm defect},q_a\right)$ & $\circ$ &  &  & $\circ$ &  & $\circ$\\
  \hline
  \end{tabular}
\end{table}
This is realized by turning on the background gauge fields $A_1$
and $B_2$ as
\begin{equation}
  A_{1,{\rm defect}}= q_\phi d\vartheta \ ,~~~
  B_{2,{\rm defect}} = 2\pi q_{a} \delta(x)\delta(y)\theta(r-\epsilon)
dx\wed dy
\ ,
\end{equation}
where $\epsilon>0$ is a regulator that is sent to zero
eventually.
It then follows that
\begin{equation}
\int A_{1,{\rm defect}}\wedge B_{2,{\rm defect}} = 4\pi^2 q \ ,~~~
q=q_{\phi}q_{a} \ .
\label{linking=1}
\end{equation}
We define this configuration as an operator
$\sfV\sfT(\cS_{\rm defect},q)$ with a 2-dimensional support
$\cS_{\rm defect}$ given by $\bR^2\ni(t,z)$ and $q$ being the
$U(1)^{[2]}$ charge.
\eqref{linking=1} is generalized to cases where
't Hooft surfaces and axionic 3-branes are linked with
each other on a slice with constant values of $(t,z)$.

The symmetry generator $U^{[2]}_{\rm CW}(\cL,\gamma)$ 
that measures the charge of
$\sfV\sfT(\cS_{\rm defect},q)$ is obtained by setting
$C_3=\gamma \delta_3(\cL)$, where
$\cL$ is a 3-dimensional surface that surrounds $\cS_{\rm defect}$.
This is verified by computing 
the correlation function of $U^{[2]}_{\rm CW}(\cL,\gamma)$ and
$\sfV\sfT(\cS_{\rm defect},q)$ 
following the same procedure as before:
\begin{equation}
  \left\langle U^{[2]}_{\rm CW}\left(\cL,\gamma\right)\sfV\sfT(\cS_{\rm defect},q) \right\rangle = e^{i\gamma q {\rm Link}(\cS_{\rm defect},\cL)} \langle \sfV\sfT(\cS_{\rm defect},q) \rangle \ .
\end{equation}

% \paragraph*{Charged object of $U^{[1]}_{\rm CW}$}

Finally, we consider 
the CW 1-form symmetry $U(1)^{[1]}$ and charged operators for it.
By definition, these have a 1-dimensional support,  
and are composed of two 't Hooft surfaces because
the CW 1-form symmetry current is given by $(da)^2/(2\pi)^2$.
A typical configuration for the charged operator is shown below:
\begin{figure}[H]
  \centering
  \includegraphics[width=5cm]{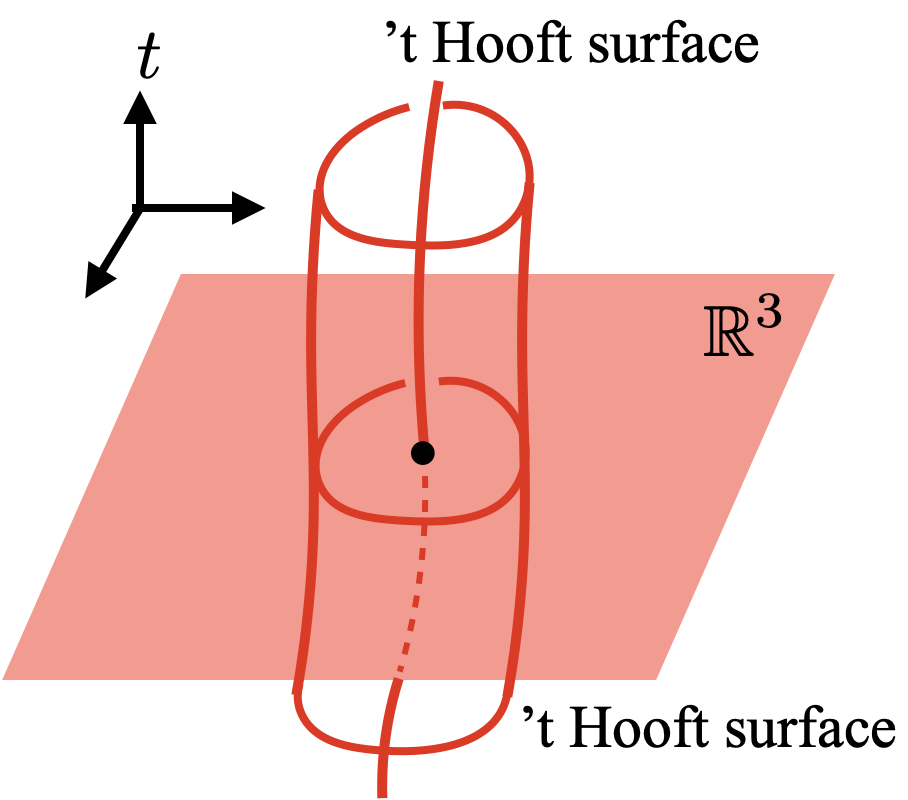}
  \caption{$\sfT\sfT(\cL^{\rm CW}_{\rm defect},q)$ 
carrying the $U^{[1]}_{\rm CW}$ charge}
  \label{1form_defect}
\end{figure}
\begin{table}[H]
  \centering
  \begin{tabular}{|c||c|c|c|c|c|c|} 
  \hline
       & $t$ & $x$ & $y$ & $r$ & $\vartheta$ & $\varphi$\\
  \hline
      $\sfT\left(\cC_{\rm defect},q\right)$& $\circ$ & $\circ$ & $\circ$ &  &  & \\
  \hline
      $\sfT\left(\cC'_{\rm defect},q'\right)$ & $\circ$ &  &  &  & $\circ$ & $\circ$\\
  \hline
  \end{tabular}
\end{table}
Here, $(r,\vartheta,\varphi)$ are the spherical coordinates of $\bR^3$,
which is transverse to $\cC_{\rm defect}$.
This configuration is realized by
\begin{equation}
  B_{2,{\rm defect}} = \frac{q}{2} \sin \vartheta d\vartheta d\phi + 2\pi q' \delta(x)\delta(y)\theta(r-\epsilon) dxdy \ ,
\end{equation}
and define the operator 
$\sfT\sfT(\cL^{\rm CW}_{\rm defect},q^{\prime\prime})$
with the 1-dimensional support $\cL^{\rm CW}_{\rm defect}$ equal to
$\bR\ni t$ and $q^{\prime\prime}$ being
the $U(1)^{[1]}$ charge evaluated from
\begin{equation}
\int B_{2,{\rm defect}}\wedge B_{2,{\rm defect}} = 8\pi^2 q^{\prime\prime} \ ,~~~
q^{\prime\prime}=qq^\prime \ .
\end{equation}
The symmetry generator $U^{[1]}_{\rm CW}\left(\cW,\xi\right)$
for measuring 
$\sfT\sfT(\cL^{\rm CW}_{\rm defect},q^{\prime\prime})$ is realized
by turning on
$B_2^{\rm CW}=\xi \delta_2(\cW)$ with
$\xi$ being the $U(1)^{[1]}$ rotation angle and $\cW$ a 4-dimensional
support that surrounds 
$\cL^{\rm CW}_{\rm defect}$.
As before, the charge of $\sfT\sfT(\cL^{\rm CW}_{\rm defect},q)$ 
results from a 't Hooft anomaly
associated with a gauge transformation for removing $B_2^{\rm CW}$:
\begin{equation}
  \left\langle U^{[1]}_{\rm CW}\left(\cW,\xi\right)\sfT\sfT(\cL^{\rm CW}_{\rm defect},q) \right\rangle = e^{i\xi q {\rm Link}(\cL^{\rm CW}_{\rm defect},\cW)} \langle \sfT\sfT(\cL^{\rm CW}_{\rm defect},q) \rangle\ .
  \label{tt}
\end{equation}

%%%%%%%%%%%%%%%%%%%%%%%%%%%%%%%%%%%%%%%%%%%%%%%%%%%%%%%%%%
\subsection{Correlation functions of symmetry generators}
\label{CF}

In this subsection, we work out the identities among correlation
function of the symmetry generators
for the purpose of understanding
the higher-group structures and their
physical interpretation in
the 6d axion-Maxwell system.
A key ingredient in this analysis is the GS transformation laws
for the CW gauge fields \eqref{B}, \eqref{C}, \eqref{D}, 
\eqref{E}.
Part of the results given below is an extension of those
obtained in \cite{Hidaka:2020iaz,Hidaka:2020izy}
for the 4d axion-Maxwell system.

\subsubsection{Correlation functions of two EoM-based symmetry generators}

We start discussing  
\begin{align}
    Z\left[\frac{2\pi n}{N}\delta_1(\cV),\frac{2\pi m}{N}\delta_2(\cW),0,0,0,0\right] &= \left\langle U_{\rm EoM}^{[0]}\left(\cV,\frac{2\pi n}{N}\right) U_{\rm EoM}^{[1]}\left(\cW,\frac{2\pi m}{N}\right)\right\rangle
\end{align}
with $A_1 = \frac{2\pi n}{N}\delta_1(\cV)$ and 
$B_2 = \frac{2\pi m}{N}\delta_2(\cW)$.
We make a gauge transformation to gauge away $B_2$:
\begin{equation}
    B_2 \rightarrow B_2 + d\Lambda_1 = 0 \ ,~~~
\Lambda_1 = \frac{2\pi m}{N}\delta_1(\Omega_\cW)
\ .
\end{equation}
Note that this gauge transformation
induces the CW gauge field $B_2^{\rm CW}$
\begin{equation}
    B_2^{\rm CW} \rightarrow B_2^{\rm CW} - \frac{N}{2\pi}A_1 \wedge \Lambda_1 = -\frac{2mn}{N}\delta_1(\cV)\wedge \delta_1(\Omega_\cW)\ ,
\end{equation}
because of \eqref{B}.
It is easy to show that no 't Hooft anomaly arises from
the gauge transformation so that
\begin{equation}
    Z\left[\frac{2\pi n}{N}\delta_1(\cV),\frac{2\pi m}{N}\delta_2(\cW),0,0,0,0\right] = Z\left[\frac{2\pi n}{N}\delta_1(\cV),0,-\frac{2\pi mn}{N}\delta_1(\cV)\wedge \delta_1(\Omega_\cW),0,0,0\right]\ .
\end{equation}
Therefore,
\begin{equation}
  \left\langle U_{\rm EoM}^{[0]}\left(\cV,\frac{2\pi n}{N}\right) U_{\rm EoM}^{[1]}\left(\cW,\frac{2\pi m}{N}\right)\right\rangle =\left\langle U_{\rm EoM}^{[0]}\left(\cV,\frac{2\pi n}{N}\right) U_{\rm CW}^{[1]}\left(\cV \cap \Omega_\cW,-\frac{2\pi mn}{N}\right)\right\rangle\ .
  \label{CF01}
\end{equation}

\begin{figure}[H]
  \centering
  \includegraphics[width=7cm]{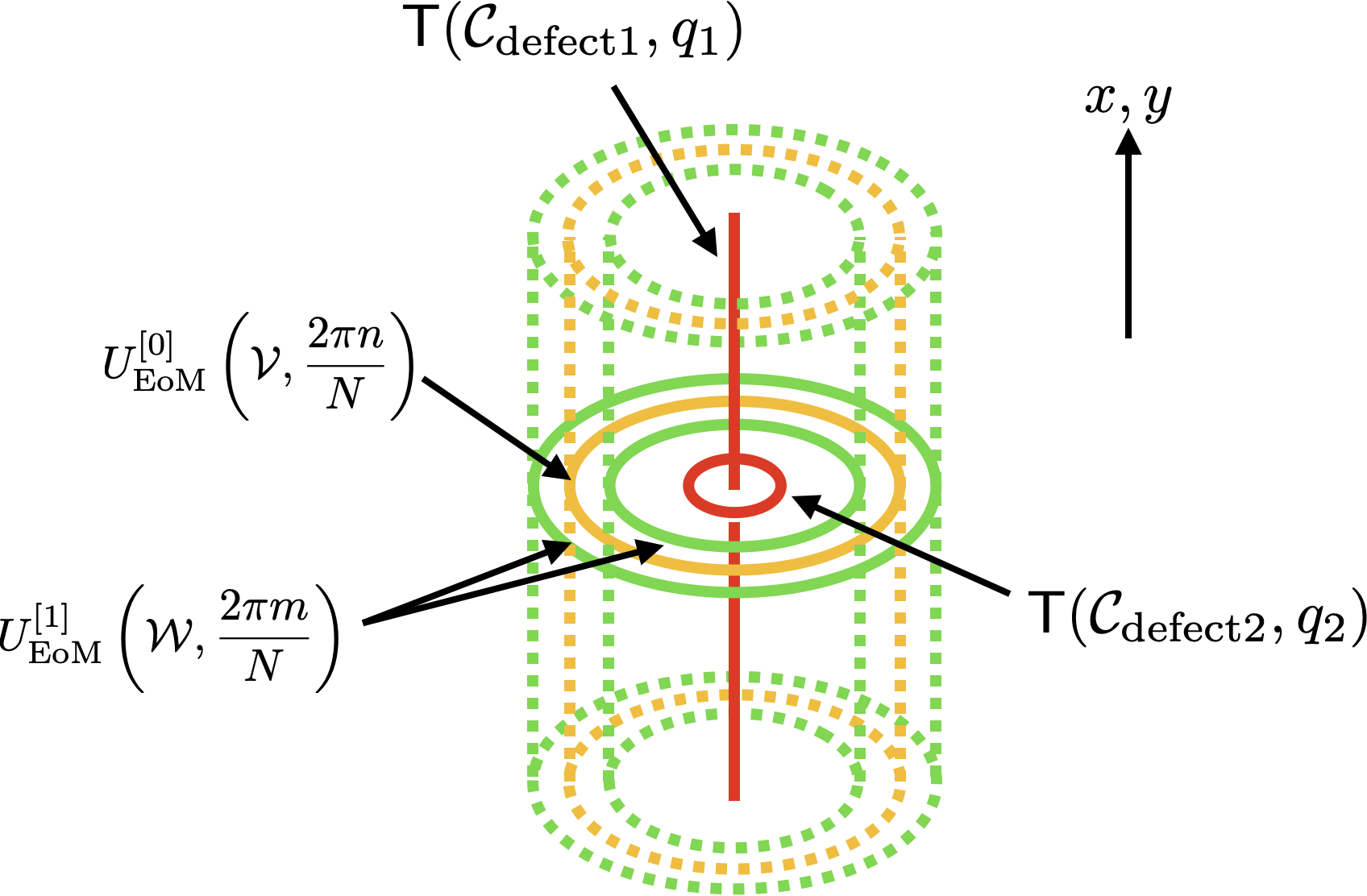}
  \caption{Witten effect on an axionic domain wall.}
  \label{wittenE}
\end{figure}

The physical meaning of this relation becomes clearer by inserting
the operator 
$\sfT\sfT(\cL^{\rm CW}_{\rm defect},q)$
into \eqref{CF01}. Using \eqref{tt}, we obtain
\begin{align}
  &\left\langle U_{\rm EoM}^{[0]}\left(\cV,\frac{2\pi n}{N}\right) U_{\rm EoM}^{[1]}\left(\cW,\frac{2\pi m}{N}\right) \sfT\sfT(\cL^{\rm CW}_{\rm defect},q) \right\rangle \nonumber\\
  =& e^{-i\frac{2\pi mnq}{N} {\rm Link}(\cL^{\rm CW}_{\rm defect},\cV\cap\Omega_\cW)} \left\langle U_{\rm EoM}^{[0]}\left(\cV,\frac{2\pi n}{N}\right)\sfT\sfT(\cL^{\rm CW}_{\rm defect},q) \right\rangle
\ .
\label{witten_CF}
\end{align}
As discussed in \cite{Hidaka:2020iaz,Hidaka:2020izy},
this can be interpreted as the Witten effect \cite{Witten:1979ey}
induced on an axion domain wall.
As a typical realization of \eqref{witten_CF},
we consider
\begin{table}[H]
  \centering
  \begin{tabular}{|c||c|c|c|c|c|c|} 
  \hline
       & $t$ & $x$ & $y$ & $r$ & $\vartheta$ & $\varphi$\\
  \hline
      $\sfT\left(\cC_{{\rm defect}1},q_1\right)$& $\circ$ & $\circ$ & $\circ$ &  &  & \\
  \hline
      $\sfT\left(\cC_{{\rm defect}2},q_2\right)$ & $\circ$ &  &  & \multicolumn{3}{c|}{$\bS^2$}\\
  \hline
      $U_{\rm EoM}^{[0]}\left(\cV,\frac{2\pi n}{N}\right)$ & $\circ$ & $\circ$ & $\circ$ & \multicolumn{3}{c|}{$\bS^2$}\\
  \hline
      $U_{\rm EoM}^{[1]}\left(\cW,\frac{2\pi m}{N}\right)$ &  & $\circ$ & $\circ$ &  \multicolumn{3}{c|}{$\bS^2$}\\
  \hline
  \end{tabular}
\end{table}
\noindent
Here, $\sfT\sfT(\cL^{\rm CW}_{\rm defect},q)$ with $q=q_1q_2$
is composed of the two 't Hooft surfaces as seen in 
Figure \ref{1form_defect}.
A plot of this configuration at $t=0$, where
$U_{\rm EoM}^{[1]}\left(\cW,\frac{2\pi m}{N}\right)$ is localized,
is shown in Figure \ref{wittenE}.
A magnetic field emanating from magnetic monopoles on the
't Hooft surfaces goes through 
%the support of
$U_{\rm EoM}^{[0]}\left(\cV,\frac{2\pi n}{N}\right)$,
which is regarded as an axionic domain wall with
the worldvolume given by $\cV$.
The phase factor appearing in \eqref{witten_CF}
implies the existence of an electric source 
induced on $\cV$, 
because
$U_{\rm EoM}^{[1]}\left(\cW,\frac{2\pi m}{N}\right)$
is designed to measure an electric flux emanating from $\cV$. 

\begin{figure}[H]
  \centering
  \includegraphics[width=7cm]{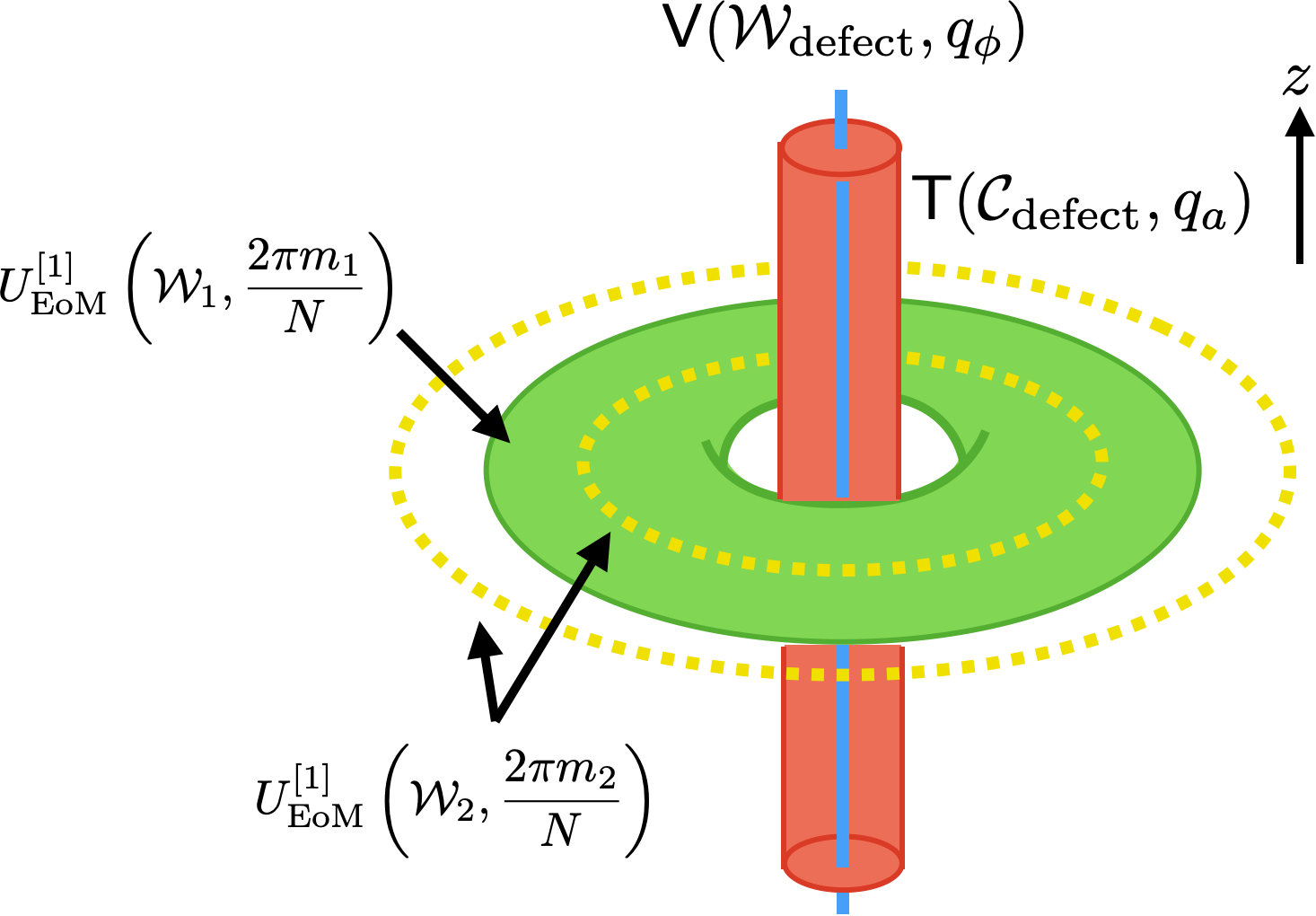}
  \caption{Anomalous Hall effect in 6d.}
  \label{hall}
\end{figure}

As a second example, we focus on the correlation function
$ \langle U_{\rm EoM}^{[1]}(\cW_1,\frac{2\pi m_1}{N})\, U_{\rm EoM}^{[1]}(\cW_2,\frac{2\pi m_2}{N})\rangle$.
This is obtained by turning on
$B_2=\frac{2\pi m_1}{N}\delta_2(\cW_1)+\frac{2\pi m_2}{N}\delta_2(\cW_2)$.
Gauging away the second term in $B_2$ to remove
$U_{\rm EoM}^{[1]}\left(\cW_2,\frac{2\pi m_1}{N}\right)$
and then using the GS transformation law \eqref{C} 
gives
\begin{equation}
  \left\langle U_{\rm EoM}^{[1]}\left(\cW_1,\frac{2\pi m_1}{N}\right) U_{\rm EoM}^{[1]}\left(\cW_2,\frac{2\pi m_2}{N}\right)\right\rangle 
= \left\langle 
U_{\rm EoM}^{[1]}\left(\cW_1,\frac{2\pi m_1}{N}\right)
U_{\rm CW}^{[2]}\left(\cW_1\cap \Omega_{\cW_2},-\frac{2\pi m_1 m_2}{N}\right)\right\rangle\ .
  \label{peiffer}
\end{equation}
We now argue that this can be interpreted as an anomalous Hall effect
in 6 dimensions. {}For this purpose, it is more convenient to
insert the operator 
$\sfV\sfT(\cS_{\rm defect},q)$
into \eqref{peiffer}.
By noting that $\sfV\sfT(\cS_{\rm defect},q)$ is charged under
$U_{\rm CW}^{[2]}$, we find
\begin{align}
  &\left\langle U_{\rm EoM}^{[1]}\left(\cW_1,\frac{2\pi m_1}{N}\right) U_{\rm EoM}^{[1]}\left(\cW_2,\frac{2\pi m_2}{N}\right)\sfV\sfT(\cS_{\rm defect},q)\right\rangle \nonumber\\
  =& e^{-i\frac{2\pi m_1m_2q}{N} {\rm Link}(\cS_{\rm defect},\cW_1\cap\Omega_{\cW_2})}\left\langle 
U_{\rm EoM}^{[1]}\left(\cW_1,\frac{2\pi m_1}{N}\right)
\sfV\sfT(\cS_{\rm defect},q)\right\rangle\ .
\label{hall_CF}
\end{align}
A typical configuration for realizing the LHS of \eqref{hall_CF}
is given below:
\begin{table}[H]
  \centering
  \begin{tabular}{|c||c|c|c|c|c|c|} 
  \hline
       & $t$ & $x$ & $y$ & $z$ & $r$ & $\vartheta$\\
  \hline
      $\sfV\left(\cW_{\rm defect},q_\phi\right)$& $\circ$ & $\circ$ & $\circ$ & $\circ$ &  & \\
  \hline
      $\sfT\left(\cC_{\rm defect},q_a\right)$ & $\circ$ &  &  & $\circ$ & \multicolumn{2}{c|}{$\bS^1$}\\
  \hline
      $U_{\rm EoM}^{[1]}\left(\cW_1,\frac{2\pi m_1}{N}\right)$ &  & $\circ$ & $\circ$ & \multicolumn{3}{c|}{$T^2$}\\
  \hline
      $U_{\rm EoM}^{[1]}\left(\cW_2,\frac{2\pi m_2}{N}\right)$ & $\circ$ & $\circ$ & $\circ$ &  & \multicolumn{2}{c|}{$\bS^1$}\\
  \hline
  \end{tabular}
\end{table}
\noindent
with a section with $t=x=y=0$ plotted in Figure \ref{hall}.
\noindent
Here, $(r,\vartheta)$ is the polar coordinates
of the 2-dimensional plane transverse to the $z$-direction.
$\cW_2$ is depicted as concentric circles that sandwich $\cW_1$.

With this setup, 
the phase factor appearing in the RHS of \eqref{hall_CF}
is identified with a magnetic flux along the
$\vartheta$-direction that is measured by
$U_{\rm EoM}^{[1]}\left(\cW_2,\frac{2\pi m}{N}\right)$.
The magnetic flux is interpreted to emanate from an electric
current induced along the $\alpha$-cycle of the 2-torus $\cW_1$.
This is a manifestation of the anomalous Hall effect in 6 dimensions.
In fact, we note that
$U_{\rm EoM}^{[1]}\left(\cW_1,\frac{2\pi m}{N}\right)$ is
realized by turning on a background electric field along the 
normal direction to $\cW_1$,
which is perpendicular to that of the induced current.

As found in \cite{Hidaka:2020iaz,Hidaka:2020izy},
the correlation function \eqref{peiffer} 
is regarded as the Peiffer lifting of a 3-group.
This implies that the 6d axion electrodynamics possesses the
3-group structure as in $d=4$.
In the next subsubsection, we make a further computation
of correlation functions to gain a stringent support
that
the 6d axion electrodynamics encodes a higher-group
structure with the 3-group realized as a substructure.

\subsubsection{Correlation functions of symmetry generators of higher ranks}
\label{corr:higherrank}

Here, we discuss correlation functions involving the symmetry generators
that are absent for $d=4$.

We first turn on 
$A_1=\frac{2\pi n}{N}\delta_1(\cV),~C_3=\alpha \delta_3(\cC)$, 
which leads to the correlation function
\begin{equation}
  \left\langle U_{\rm EoM}^{[0]}\left(\cV,\frac{2\pi n}{N}\right) U_{\rm CW}^{[2]}\left(\cC,\alpha\right)\right\rangle.
\end{equation}
By gauging away $A_1$ and using the GS transformation law \eqref{D},
we find
\begin{equation}
  \left\langle U_{\rm EoM}^{[0]}\left(\cV,\frac{2\pi n}{N}\right) U_{\rm CW}^{[2]}\left(\cC,\alpha\right)\right\rangle =\left\langle U_{\rm CW}^{[2]}\left(\cC,\alpha\right) U_{\rm CW}^{[3]}\left(\cV\cap \cC,\frac{\alpha n}{N}\right)\right\rangle\ .
\end{equation}
As another example where the action of a symmetry generator of a lower
rank
gives rise to $U^{[3]}_{\rm CW}$,
we find
\begin{equation}
  \left\langle U_{\rm EoM}^{[1]}\left(\cW,\frac{2\pi m}{N}\right) U_{\rm CW}^{[1]}\left(\cW^{\rm CW},\beta\right)\right\rangle =\left\langle U_{\rm CW}^{[1]}\left(\cW^{\rm CW},\beta\right) U_{\rm CW}^{[3]}\left(\cW\cap \cW^{\rm CW},\frac{\beta m}{N}\right)\right\rangle\ .
\end{equation}
Here, the LHS is defined by the partition function
in the presence of 
$B_2=\frac{2\pi m}{N}\delta_1(\cW),\,B_2^{\rm CW}=\beta \delta_1(\cW^{\rm CW})$,
while the RHS is obtained by gauging away $B_2$.

It is verified that the action of $U_{\rm EoM}^{[1]}$ on
$U^{[2]}_{\rm CW}$ gives rise to the 4-form symmetry
generator:
\begin{equation}
  \left\langle U_{\rm EoM}^{[1]}\left(\cW,\frac{2\pi m}{N}\right) U_{\rm CW}^{[2]}\left(\cC,\gamma\right)\right\rangle =\left\langle U_{\rm CW}^{[2]}\left(\cC,\gamma\right) U_{\rm CW}^{[4]}\left(\cW\cap \cC,\frac{\gamma n}{N}\right)\right\rangle\ .
\end{equation}

We next discuss correlation functions of three
symmetry generators of lower ranks.
As a first example, we turn on  
$A_1=\frac{2\pi n}{N}\delta_1(\cV),\,B_2=\frac{2\pi m_1}{N}\delta_2(\cW_1)+\frac{2\pi m_2}{N}\delta_2(\cW_2)$ to define the
3-point function of $U_{\rm EoM}^{[0]}$ and $U_{\rm EoM}^{[1]}$.
By gauging away both $A_1$ and $B_2$, the 3-point function becomes
a correlation function involving $U_{\rm CW}^{[3]}$:
\begin{align}
  &\Big\langle U_{\rm EoM}^{[0]}(\cV,\frac{2\pi n}{N})\, U_{\rm EoM}^{[1]}(\cW_1,\frac{2\pi m_1}{N})\, U_{\rm EoM}^{[1]}(\cW_2,\frac{2\pi m_2}{N}) \Big\rangle \nonumber\\
  &= \Big\langle U_{\rm CW}^{[1]}(\cV\cap \Omega_{\cW_1},\frac{2\pi n m_1}{N})
\,U_{\rm CW}^{[1]}(\cV \cap \Omega_{\cW_2},-\frac{2\pi n m_2}{N})
\nonumber\\   
&~~\times U_{\rm CW}^{[2]}(\cW_1\cap \Omega_{\cW_2},-\frac{2\pi m_1 m_2}{N}) 
\,U_{\rm CW}^{[3]}(\cW_1\cap \cW_2 \cap \Omega_{\cV},-\frac{4\pi n m_1 m_2}{N^2}) \Big\rangle\ .
\end{align}% 
{}Furthermore,
turning on 
$B_2=\frac{2\pi m_1}{N}\delta_2(\cW_1)+\frac{2\pi m_2}{N}\delta_2(\cW_2)+\frac{2\pi m_3}{N}\delta_2(\cW_3)$
and then gauging it away gives
a correlation function involving $U_{\rm CW}^{[4]}$:
\begin{align}
  &\Big\langle U_{\rm EoM}^{[1]}(\cW_1,\frac{2\pi m_1}{N})\, U_{\rm EoM}^{[1]}(\cW_2,\frac{2\pi m_2}{N}) \,U_{\rm EoM}^{[1]}(\cW_3,\frac{2\pi m_3}{N}) \Big\rangle \nonumber\\
  &= \Big\langle U_{\rm CW}^{[2]}(\cW_1\cap \Omega_{\cW_2},-\frac{2\pi m_1 m_2}{N})\,U_{\rm CW}^{[2]}(\cW_2\cap \Omega_{\cW_3},-\frac{2\pi m_2 m_3}{N}) 
  \,U_{\rm CW}^{[2]}(\cW_3\cap \Omega_{\cW_1},-\frac{2\pi m_3 m_1}{N})
\nonumber\\
&~~\times U_{\rm CW}^{[4]}(\cW_1\cap \cW_2 \cap \Omega_{\cW_3},-\frac{4\pi m_1 m_2 m_3}{N^2}) 
\Big\rangle\ .
\label{EoM3}
\end{align}

These results are regarded as a manifestation of 
the algebraic structures that are peculiar to 
the 6d axion-Maxwell theory.
In particular, it follows from the last two computations that
the higher-group structure in this theory should be
equipped with a trinary operation
among three symmetry generators.

\section{Conclusion and Discussion}
\label{conc} 

In this paper, we discuss higher-dimensional axion electrodynamics
for the purpose of exploring a higher-group structure encoded in it
by generalizing the results in
\cite{Hidaka:2020iaz,Hidaka:2020izy}.

We first discuss how the operator-valued ambiguities that arise
from gauging EoM-based global symmetries are canceled.
This is achieved by gauging $(2n-2)$ Chern-Weil symmetries
simultaneously.
It is crucial that the CW gauge fields make a Green-Schwarz 
transformation under the EoM-based symmetry transformation
in order to guarantee gauge invariance of the resultant theory.

The main focus of this paper is on the 6d axion-Maxwell system.
We give the explicit form of the GS transformation of the 
four CW gauge fields. We also determine the 't Hooft anomaly
due to an ambiguity of how to extend the system
to a 7d spacetime.
We next compute correlation functions of the symmetry generators
by employing the fact that any configuration of the
symmetry generators and charged operators
is constructed by turning on the background gauge fields
appropriately. The correlation functions of two configurations 
are equal to each other up to a 't Hooft anomaly if
they are mapped to each another by a gauge transformation.
On top of correlation functions that have been obtained
already in \cite{Hidaka:2020iaz,Hidaka:2020izy},
we work out a new class of correlation functions that
are peculiar to the $d=6$ case.
These results suggest that 
 the 6d axion-Maxwell system possesses 
a higher-group structure such that
the 3-group structure found in the 4d axion-Maxwell system 
is encoded as a substructure. 
{}Furthermore, it is natural to expect that 
the possible higher-group structure should admit a trinary operation,
an algebraic structure involving three symmetry generators,
as discussed in section \ref{corr:higherrank}.

More generally, the axion-Maxwell system in $d=2n$ dimensions 
is expected to possess a higher-group structure with
a substructure identical to that of the $d=2n-2$ axion-Maxwell
system.
This is because all the CW gauge field strengths for the $d=2n-2$ case
are included in those for the $d=2n$ case.
{}Furthermore, the higher-group structure for the $d=2n$ case, 
if exists, should admit an $n$-ary operation among $n$ symmetry
generators. To see this, we note that 
the $d=2n$ axion-Maxwell system has
the $(2n-2)$-form symmetry with the CW current $d\phi/(2\pi)$,
and it couples to a $(2n-1)$-form CW gauge field.
The gauge invariant $2n$-form field strength 
contains a term proportional to $(B_2)^n$.
This implies that two correlation functions, one with a
single insertion of the symmetry generator $U_{\rm CW}^{[2n-2]}$
and the other with $n$ insertions of $U_{\rm EoM}^{[1]}$,
are related with each other as found in \eqref{EoM3} for $n=3$.

In this paper, we have not attempted to formulate rigorously 
the mathematical structure of the higher-group symmetry that underlies 
the higher-dimensional
axion-Maxwell systems. We leave it for future work.

\subsection*{Acknowledgements}
RY is supported by JSPS KAKENHI Grants No.~JP21J00480, JP21K13928.

\appendix
\section{Alternative method of computing correlation functions}
\label{alt}

In section \ref{CF}, correlation functions are computed
using a network of background gauge fields
and gauge transformations acting on it.
Here, we review an alternative way that is developed in
\cite{Hidaka:2020iaz} for the 4d axion-Maxwell system.

Let $S[\phi,a]$ be the action \eqref{action}.
Shifting the axion and the Maxwell field
by the background gauge fields $\Phi_0$ and
$\Pi_1$, respectively, we find
\begin{align}
  S[\phi,a] + \int_{\cM_{2n}} dj^{[0]}_{\rm EoM}\wedge \Phi_0 &= S[\phi-\Phi_0,a] + \frac{1}{2}\int_{\cM_{2n}} d\Phi_0 \wedge \star d\Phi_0 
\ ,\label{action-jphie}\\
  S[\phi,a]+ \int_{\cM_{2n}} dj^{[1]}_{\rm EoM}\wedge \Pi_1 &= S[\phi,a+\Pi_1]+ \frac{N}{(2\pi)^n}\sum_{r=2}^n \frac{1}{(n-r)! r!}d\phi (da)^{n-r} \wedge (d\Pi_1)^{r-1} \wedge \Pi_1 \nonumber\\
  &~~~ + \frac{1}{2}\int_{\cM_{2n}} d\Pi_1 \wedge \star d\Pi_1 \ .
\label{action-jae}
\end{align}
These results play a key role in the computations
made below.

As a sample computation, we discuss
the correlation function of  
$U_{\rm EoM}^{[0]}(\cV,\frac{2\pi n}{N})$
and
$U_{\rm CW}^{[1]}(\cW,\frac{2\pi m}{N})$ for the $n=3$ case.
Noting that $U_{\rm EoM}^{[0]}(\cV,\frac{2\pi n}{N})$ is rewritten
as
\begin{align}
 U_{\rm EoM}^{[0]}(\cV,\frac{2\pi n}{N})=\exp(\frac{2\pi}{N}\int_\cV j_{\rm EoM}^{[0]})=
\exp(\frac{2\pi}{N}\int_{\cM_6} dj_{\rm EoM}^{[0]}\wedge \delta_0(\Omega_\cV)) \ ,
\end{align}
it follows from \eqref{action-jphie} with
$\Phi_0=(2\pi/N)\,\delta_0(\Omega_\cV)$
that
\begin{equation}
  \left\langle U_{\rm EoM}^{[0]}\left(\cV,\frac{2\pi n}{N}\right) U_{\rm EoM}^{[1]}\left(\cW,\frac{2\pi m}{N}\right)\right\rangle = {\cal N} \int {\cal D}[\phi,a] e^{iS[\phi-\frac{2\pi n}{N}\delta_0(\Omega_\cV),a] + \frac{2\pi i m}{N}\int_\cW j^{[1]}_{\rm EoM}}\ .
\end{equation}
By shifting 
$\phi\rightarrow\phi' = \phi-\frac{2\pi n}{N}\delta_0(\Omega_\cV)$,
we obtain
\begin{align}
  \left\langle U_{\rm EoM}^{[0]}\left(\cV,\frac{2\pi n}{N}\right) U_{\rm EoM}^{[1]}\left(\cW,\frac{2\pi m}{N}\right)\right\rangle  =& {\cal N} \int {\cal D}[\phi',a] e^{iS[\phi',a] + \frac{2\pi i m}{N}\int_\cW j^{[1]}_{\rm EoM}-\frac{inm}{4\pi N}\int_\cW da\wedge da \delta_0(\Omega_\cV)} \nonumber\\
  =& \left\langle U_{\rm EoM}^{[1]}\left(\cW,\frac{2\pi m}{N}\right)U_{\rm CW}^{[1]}\left(\cW\cap \Omega_\cV,-\frac{2\pi nm}{N}\right) \right\rangle \ ,
\end{align}
because $j_{\rm EoM}^{[1]}$ gets shifted under the shift.
This coincides with \eqref{CF01}.

The rest of the correlations computed in this paper
can be reproduced following the the same way as discussed
in this appendix.

% \bibliography{yokokura}
\providecommand{\href}[2]{#2}\begingroup\raggedright\endgroup

\end{document}